# Performance of FPGA controller in ISAC-1 accelerator chain

K. Fong, X. Fu, Q.W. Zheng, T. Au, R. Leewe, TRIUMF, V6T2A3, Vancouver, Canada

*Abstract*

The LLRF of five of TRIUMF's ISAC-1 accelerator cavities have been replaced by 3 similar FPGA based system with different operating frequencies. These LLRF use internal digital phase locked loops for frequency generation and synchronization, feedback control using Amplitude/Phase regulations. These FPGAs also have internal stepper motor controller for resonance control. Various modes of resonance control are possible, including phase comparison and minimum seeking slide-mode control. Operational performances including frequency generation and synchronization, amplitude and phase noises, tuning speeds, compatibility to original remote controls, are reported.

*Introduction*

The LLRF of the original ISAC-1 accelerator RF system was VXI-based commissioned 2000. After 23 years of operations, spare parts for the DSP LLRF are getting difficult to obtain. The most failed parts are however the VXI mainframe and its built-in power supply. For these reasons, TRIUMF has embarked on replacing the LLRF controllers for ISAC-1. The controllers that were replaced in 2023 are the prebuncher, 2 Drift Tube Linacs and 2 bunchers, namely DTL4, DTL5, HEBT11 and HEBT35. The operating frequencies for the prebuncher are 11.786 MHz, 23.573 MHz and 35.36 MHz. The bunchers are 11.786 MHz, 35.36 MHz and 106.08 MHz for both DTLS, with the RF power ranges from 1.5 kW and 13 kW for the 2 bunchers to more than 20 kW for the DTLs. The new system has a core FPGA, designed by TRIUMF. It is mounted within a NIM module, and controlled via an USB/HID interface[1]. Each FPGA can control up to 3 different phase-locked frequencies, providing amplitude and phase regulations, as well as stepping motor control for tuner operation. Amplitude and phase regulations are available for each channel. Table 1 gives a summary of the difference between the old VXI-based system and the NIM bin system.

*USB/HID interface*

The FPGA communicates to the external world uses USB/HID protocol. HID is a device class to supplement the USB interface. It is very compact but extensible and robust. Each report is 48-bits long, for transmitting commands and receiving status from the FPGA. A USB/HID based system offers many advantages over a much older VXI based system. No additional hardware or bus is required. Any modern computer offers multiple USB ports, device drivers are available in all operating systems. These operating systems also provide extensive diagnostics and recovery for signal interruptions. Due to the use of Python interpreter in the FPGA, the timing of each report is around 3 ms, but is sufficient for the purpose of supervisory control.

*LLRF controller Algorithm*

The FPGA is configured to use amplitude/phase feedback regulation as opposed to the more commonly used I/Q feedback regulation. I/Q feedback is the preferred choice when the performance of analogue phase detector and modulator was poor, but with the advances of high speed digital circuitry and algorithm such as CORDIC, amplitude/phase feedback can have the same precision as I/Q feedback. In addition, amplitude/phase feedback avoids cross talk in I/Q feedback. This crosstalk can induce 'AM-PM Conversion Instabilty'.[2] whereas amplitude/phase system is immune to this kind of instability. Fig. 1 shows the block diagram of an I/Q controller, while Fig. 2 shows that with the addition of a CORDIC and a NCO, they transform the cartesian coordinates to polar coordinates to form an Amplitude/Phase controller. The signal flow chart of an I/Q feedback controller and an Amplitude/Phase controller are shown in Fig.3 and Fig.4 respectively[3]. In the case of tuner is at resonant, it can be seen that there is no crosstalk between the Amplitude channel and the Phase channel, while in an I/Q controller, the additional of the loop delay will introduce cross-talks. In normal operating conditions, all these cross-talks are eliminated so there should not be any performance differences between I/Q and Amplitude/Phase feedback controller. To verify this, we have set up a measurement system as shown in Fig. 5 which can switch rapidly between the 2 systems and proceed to measure their respective noises.

| VXI based controller | NIM based controller |
|---|---|
| 2 DSPs | 1 FPGA |
| VXI interface | USB-HID interface |
|  | Hot pluggable |
| 1 RF channel | 3 RF channels |
| Single frequency | Multi-frequencies |
| Base band | 31.6 MHz IF, 75 MHz LO |
| 200 kHz sampling | 250 MHz sampling |
| I/Q control | Amplitude/Phase control |
| External stepper controller | Internal stepper controller |
|  | Feedforward Ramping |

Table 1. Comparison between VXI based controller and NIM based controller.

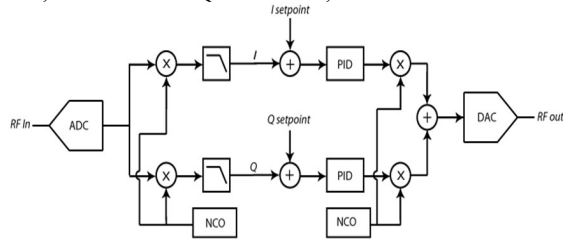

Figure 1. Block diagram of an I/Q controller

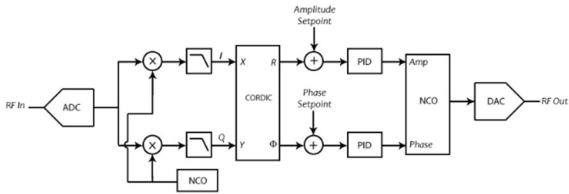

Figure 2. Block diagram of an Amplitude/Phase controller realized in digital circuitry.

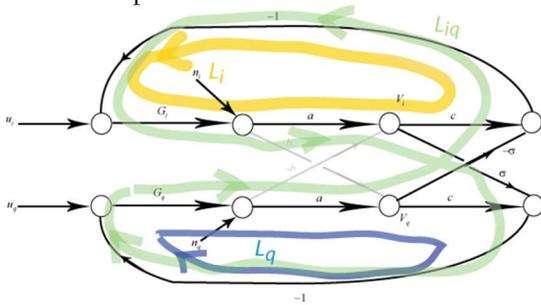

Figure 3. Signal flow chart of an I/Q feedback controller showing a cross-talk path due to bad loop delay.

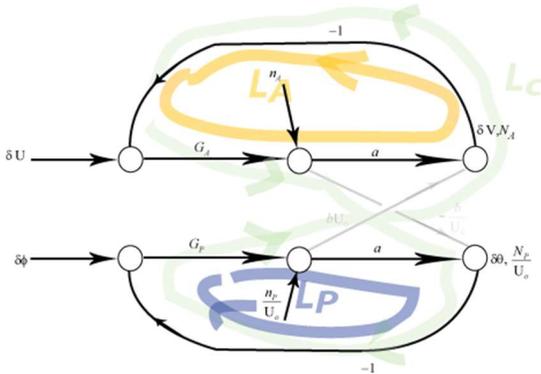

Figure 4. Signal flow chart of an Amplitude/Phase feedback controller showing the absence of cross talk due from loop delay.

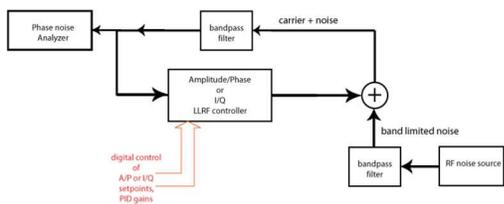

Figure 5. Measurement setup to compare performances of I/Q vs Amplitude/Phase controllers.

A rf noise source is used to inject random noise inside the feedback loop. The measurement results are shown in Fig.6 and Fig. 7. These clearly show that there is no degradation in noise performance between the I/Q controller and the Amplitude/Phase controller where a CORDIC is added. Finally, a comparison under actual operating conditions is make by measuring the AM and Phase noise of

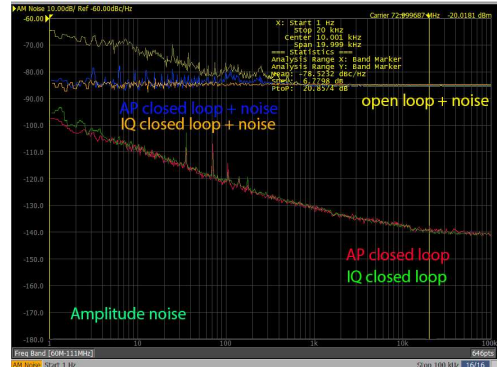

Figure 6. Comparison of AM noises in I/Q and Amplitude/Phase controllers.

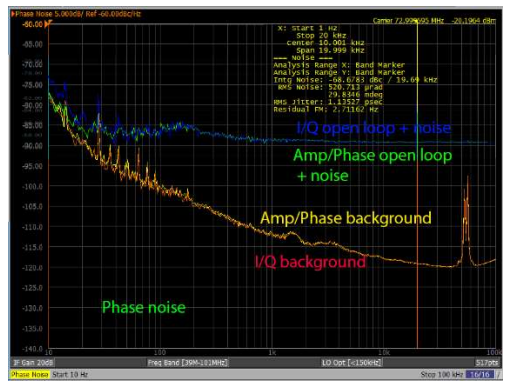

Figure 7. Comparison of Phase noise in I/Q and Amplitude feedback controllers.

DTL3, which still uses the VXI-base system, to DTL5, which has been converted to Amplitude/Phase system. These results are shown in Fig.8 for AM noise and Fig.9 for Phase noise. Again the results show there is negligible difference in both the AM and Phase noise. There are 2 peak for the new system at 30 kHz and 40 kHz, which are due to higher than necessary gains of the PLL. The gains will be reduced in the next available shutdown period.

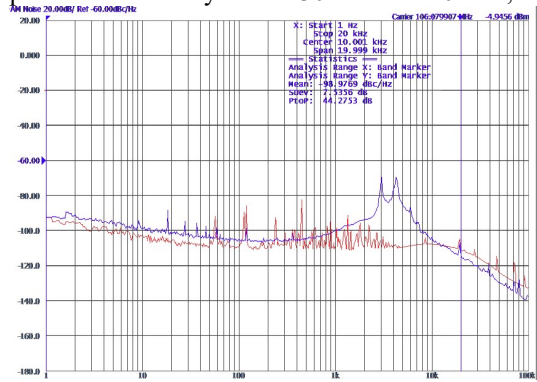

Figure 8. AM noise of DTL3 (red curve) and DTL5 (blue curve).

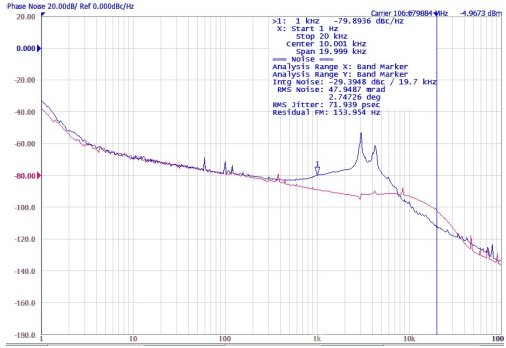

Figure 9. Phase noise of DTL3 (red curve) and DTL5 (blue curve).

*Automatic Power-up Sequence*

The new system combines 3 tuning algorithms: Positional alignment, phase comparison and *s*liding *m*ode *ex*tremum *s*eeking. Since each of these algorithms has its advantages and its disadvantages, they are operated in sequence to utilize their advantages and avoiding their disadvantage. The main disadvantage of the positional alignment and phase comparison is that both required predetermined setpoints that is temperature dependent. The sequence in shown in Fig. 10, where at first positional alignment is used to move the tuner to a pre-set position where rf can be established in the cavity. This is the region coloured in blue. When rf is established and in CW, phase comparison is employed to bring the reflected power down to a level where the sliding mode is most efficient and finally sliding mode is deployed to keep the reflected power minimum, irrespective of the cavity's and amplifier's tuning dependence on temperature.

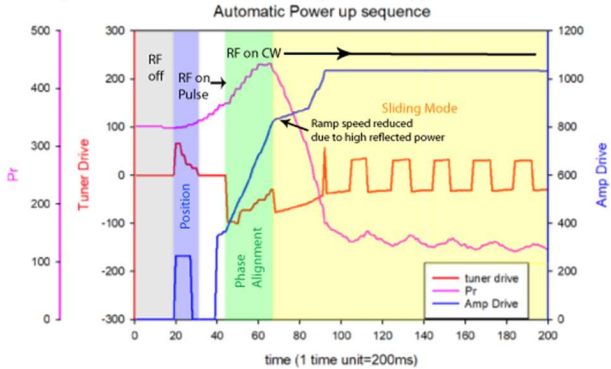

Figure 10. Sequences in automatic tuning on powering -up.

*Performance of Sliding Mode*

Because the stepper controller has been incorporated within the FPGA, it becomes possible to have more complex algorithms when using the sliding mode extremum seeking algorithms as well as incorporating phase comparison method for pretuning. The details of sliding mode equation is given as[4,5,6]

$$\dot{\theta} = k_0 \, \text{sgn}\left[\sin\left(\frac{\pi s}{\varepsilon}\right)\right] \quad (1)$$

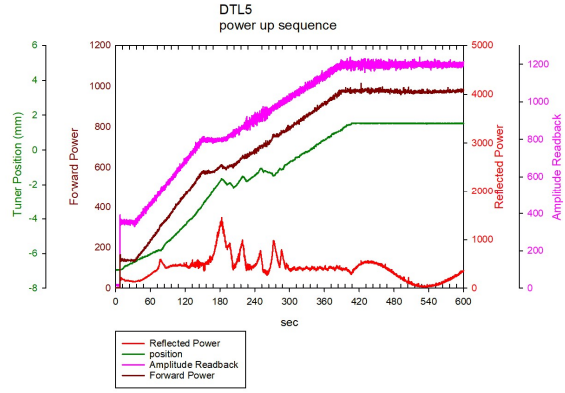

Figure 11. Powering up sequence in DTL5

With $s(t) = P_r + \rho t$ (2)

$k_0$ is the maximum speed, $\rho, \varepsilon$ are sliding mode tuning parameters, and the function to be minimized is $P_r$, the reflected power, whereas $\theta$ is the distance to minimum. The way the sliding mode extremum seeking algorithm works is clear by expressing the differential form of Eq. 2, where

$$\frac{dP_r}{dt} + \rho \approx < 0 \quad (3)$$

and since $k_0 = \frac{d\theta}{dt}$

$$\frac{dP_r}{d\theta} k_0 + \rho \approx < 0 \quad (4)$$

When this condition is satisfied, the tuner will move in a direction where $P_r$ is being reduced. The integral form as in Eq. 2 is used because the derivative in Eq. 3 tends to generate a lots of high frequency noise. Sliding mode works best when the tuning is less than 1 bandwidth from the resonance frequency, where the slope of $P_r$ vs $\theta$ is steepest. The selection of $\rho$ depends on many factors such cavity bandwidth, tuning motor speed $k_0$, reflected power $P_r$ as well as sampling time. Although the sliding mode can be operated within a wide range of $\rho$ and $\varepsilon$, an automatic sequence is used to estimate the best $\rho$ value given the maximum tuner speed, To determine the numerical value of $\varepsilon$, given Eq. 2, when the tuner is moving in the correct direction, $s \approx 0$, but if the tuner is moving in the wrong direction, $s = 2\rho t$. Therefore, by choosing $\varepsilon$ such that $t < 0.5\,\varepsilon/\rho$, the tuner will only move in the wrong direction for at most time *t*. Finally, the maximum speed $k_0$ can be set to be the same as the maximum speeds used in the position pretuning and phase alignment modes, where the speeds are set such that slight under-damping occurs.

## Chatter Reduction using Surface Skipping

As with all extremum seeking algorithms, chatter presented in the controller can degrade its performance and cause unnecessary mechanical wear. Various methods has been proposed to reduce this chatter.[7,8] We use a method which is similar to the "boundary layer method[9], by observing the rate at which the minimizing function approaches the sliding surface, it is possible to determine whether a change in direction is necessary, thereby reducing the amount of chatter throughout the minimum seeking process. For this we examine when the inequality in Eq. 3 is valid, , we can add $t = t + \Delta t$ to increase $s$ when $s < 0.1$. We call this "switching surface skipping", as it behaves very much like a stone skipping on the surface of water. Fig 12 shows a simulation of both the reflected power and the tuner position, with or without surface skipping. As shown in the figure, surface skipping can reduce the convergence time by a factor of 1/2. Fig. 13 show the behaviour of DTL5 during automatic powering up sequence. The system went through position pre-tuning mode very quickly. While the tuner is in phase compensation mode, as the forward power is ramping up, internal heating detune the cavity causing the reflected power to rise, until the system enters the sliding mode at around 9 seconds. Then the reflected power stayed down for the rest of the ramping process.

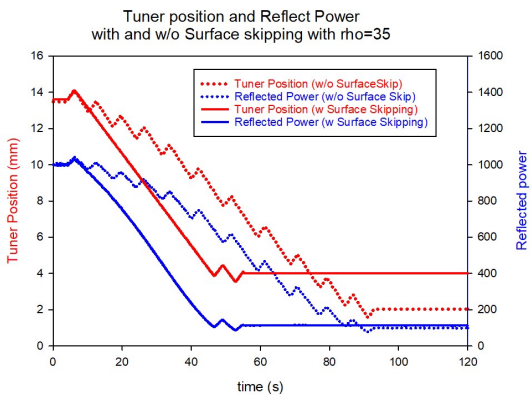

Figure 12. Chatter reduction using Surface Skipping

## Tuning aids in Sliding Mode

The Tuner local GUI is shown in Fig. 14 showing prominent parameters of the tuning control. The selection of the sliding mode parameters $\varepsilon$, $\rho$, can affect the stability of the sliding mode, and the selection of their values are not obvious. For these reasons two visual tuning aids are available in addition to an automatic $\rho$ estimation routine. The automatic $\rho$ estimation routine moves the tuner back and forth at around the steepest region of reflected power vs tuner position to calculate $\rho$. The visual aid are an "Attitude Indicator" spinning wheel, displaying $\dfrac{s\pi}{\varepsilon}$, and a moving tape display $s$. The optimal parameters will be when both the moving tape and the spinning wheel remains stationary.

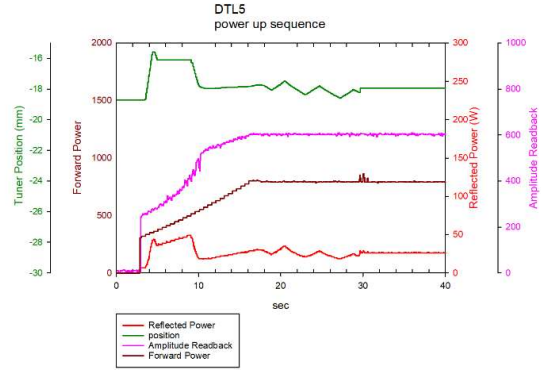

Figure 13. Automatic Powering up sequence of DTL5.

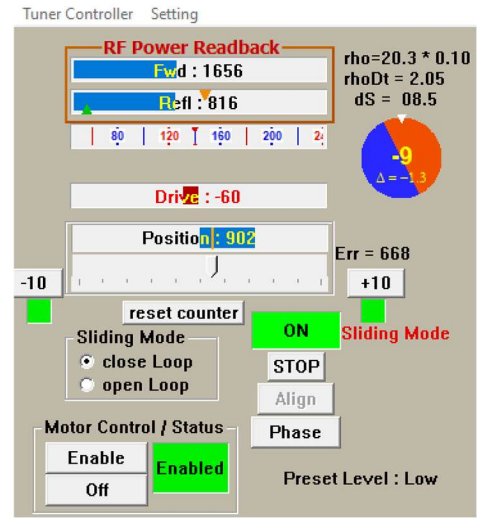

Figure 14. Tuner Window showing tuning aids

## Long Term Performance

The long term performance over 3 days of DTL5 is plotted in Fig. 15. It is clear from the figure that the diurnal ambient temperature fluctuation is about 5 ºC. The tuner position shows a corresponding fluctuation as the ambient temperature. The forward power shows a smaller fluctuation, but the reflected power remains unaffected.

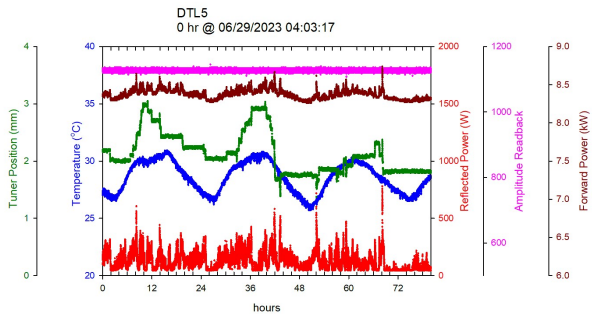

Figure 15. Long term performance of Tuning response for DTL5

## CONCLUSION

The USB/HID FPGA controllers have been in operation since April, 2023 and have operated satisfactory. Amplitude/Phase regulation provides similar performance as I/Q regulation, without the need to control the loop phase. The position preset, phase alignment and sliding mode controllers are used in the new ISAC-1 resonance control. Based on each system's strength and weakness, they are used at different stages of powering up. The tuning aids for sliding mode operation have enabled selection of tuning parameter to be easier. During the sliding mode operation, the amount of chattering can be suppressed by the "switch surface skipping" method.